\documentclass[english,aip, rsi, twocolumn]{revtex4}
\usepackage[T1]{fontenc}
\usepackage{babel}
\usepackage{amstext}
\usepackage{graphicx}
\usepackage[unicode=true,
 bookmarks=true,bookmarksnumbered=false,bookmarksopen=false,
 breaklinks=false,pdfborder={0 0 1},backref=false,colorlinks=false]
 {hyperref}
\usepackage{color}
\makeatletter
\@ifundefined{textcolor}{}
{%
 \definecolor{BLACK}{gray}{0}
 \definecolor{WHITE}{gray}{1}
 \definecolor{RED}{rgb}{1,0,0}
 \definecolor{GREEN}{rgb}{0,1,0}
 \definecolor{BLUE}{rgb}{0,0,1}
 \definecolor{CYAN}{cmyk}{1,0,0,0}
 \definecolor{MAGENTA}{cmyk}{0,1,0,0}
 \definecolor{YELLOW}{cmyk}{0,0,1,0}
}

\makeatother

\begin{document}

\newcommand{\change}[2][]{{#2}}
\preprint{This line only printed with preprint option}

\title{A compact, robust, and transportable ultra-stable laser with a fractional
frequency instability of $1\times10^{-15}$}

\author{Qun-Feng Chen}

\author{Alexander Nevsky}

\author{Marco Cardace}

\author{Stephan Schiller}

\email{Step.Schiller@uni-duesseldorf.de}

\affiliation{Institut f\"{u}r Experimentalphysik, Heinrich-Heine-Universit\"{a}t
D\"{u}sseldorf, 40225 D\"{u}sseldorf, Germany}

\author{Thomas Legero} \author{Sebastian H\"{a}fner} \author{Andre Uhde} \author{Uwe Sterr}

\affiliation{Physikalisch-Technische Bundesanstalt (PTB), Bundesallee 100, 38116
Braunschweig, Germany }
\begin{abstract}
We present a compact and robust transportable ultra-stable laser system
with minimum fractional frequency instability of $1\times10^{-15}$
at integration times between 1 to 10 s. The system was conceived as
a prototype of a subsystem of a microwave-optical local oscillator
to be used on the satellite mission STE-QUEST (Space-Time Explorer
and QUantum Equivalence Principle Space Test, http://sci.esa.int/ste-quest/).
It was therefore designed to be compact, to sustain accelerations
occurring during rocket launch, to exhibit low vibration sensitivity,
and to reach a low frequency instability. Overall dimensions of the
optical system are $40\textrm{ cm}\times20\textrm{ cm}\times30\textrm{ cm}$.
The acceleration sensitivities of the optical frequency in the three
directions were measured to be $1.7\times10^{-11}/g$, $8.0\times10^{-11}/g$,
and $3.9\times10^{-10}/g$, and the absolute frequency instability
was determined via a three-cornered hat measurement. The design is
also appropriate and useful for terrestrial applications.
\end{abstract}
\maketitle

\subsection{Introduction}
\change[Because of the free-fall conditions, a possibly varying distance to
Earth and varying speed, of satellites on selected orbits in Space,
Space are]{Satellites on selected orbits in Space is possible to experience a
long-duration of free-fall condition, a varying distance to Earth (or to other
planets), a varying speed, or a varying line of sight to the satellite. These
conditions make Space} an ideal place to carry out certain precision
experiments, in particular for testing fundamental notions of space and time
\cite{ashby2009measurement,schiller2009einstein,wolf2009quantum,thespacetime,schiller2012thespace,altschul2014quantum}.
For example, atomic clocks, including optical clocks, are proposed to be
operated in Space for testing the gravitational time dilation or the Shapiro
effect. In such missions, ultra-stable lasers are foreseen to be used as the
local oscillators to interrogate the atomic transitions. 

A laser with ultra-stable frequency is obtained by locking the laser
frequency to the resonance of a high-finesse cavity, which makes the
fractional frequency instability $\Delta\nu/\nu$ of the laser equal
to the cavity's fractional length instability $\Delta L/L$. This
latter instability is in practice limited by several factors: the
Brownian thermal noise of the cavity, the temperature \change{in}stability, and
the mechanical \change{in}stability (influenced by environmental seismic and
acoustic accelerations) of the setup. Usually, in the laboratory,
in order to reduce the influence of the unavoidably present acceleration
noise to the cavity, the cavity is laid on optimized supporting points
\cite{nazarova2006vibrationinsensitive,webster2007vibration,millo2009ultrastable},
which makes the whole apparatus not easily transportable. For transportable
setups, the cavities need to be mechanically locked during transportation
or need to be squeezed in a supporting structure \cite{vogt2011demonstration}.
In this case deformations of the supporting structure also deform
the cavity and increase its acceleration sensitivities. In order to
reduce this effect, cavities are designed with special, optimized
shapes and squeezed at a particular angle \cite{webster2011forceinsensitive,leibrandt2011spherical}.
Using such squeeze-insensitive designs, cavities can be operated in
non-laboratory environments \cite{leibrandt2011fieldtest}. However,
these designs employ approximately cubic cavity shapes which implies
that the cavity volume increases by a factor $L^{3}$, if the cavity
length $L$ is increased, which is necessary e.g. in order to reduce
its thermal noise. Moreover, it is unclear whether the mountings developed
so far are robust enough to withstand a rocket launch, a necessary
condition for use of the cavity in space. Cavity designs aimed at
satisfying this requirement have been developed \cite{folkner2010laserfrequency,argence2012prototype}
which comprise a complex mechanical layout which rigidly mounts the
cavity and at the same time decouples it from acceleration induced
deformations of the supporting structure. 

In the present publication, we report on another candidate design
for a space cavity, which uses independent separate rigid mountings
for the each of the three translational degrees of freedom, each acting
in the corresponding symmetry plane \cite{uwe2012frequenzstabilisierungsvorrichtung}.
It is relatively simple in its implementation, and therefore should
be of interest also for easily transportable terrestrial applications.
Moreover, our design is applicable for long and slim cavities, which
are thus relatively compact and lightweight. The system we present
was developed for use together with a Nd:YAG laser, since commercial
space-qualified Nd:YAG lasers with low free-running linewidth and
small free-running absolute frequency drift are available (and have
flown in space). 

Specifically, within the concept of the STE-QUEST mission\cite{thespacetime},
the purpose of the frequency-stabilized Nd:YAG laser is to serve as
an optical local oscillator. An erbium-doped fiber laser frequency
comb will be used to convert the ultra-stable optical frequency into
an ultra-stable microwave signal for interrogation of a cold cesium
atomic clock. This microwave is specified with orders of magnitude
lower phase noise and short-term frequency instability than the best
space-qualified quartz oscillators.

\begin{figure}
\includegraphics[width=8cm]{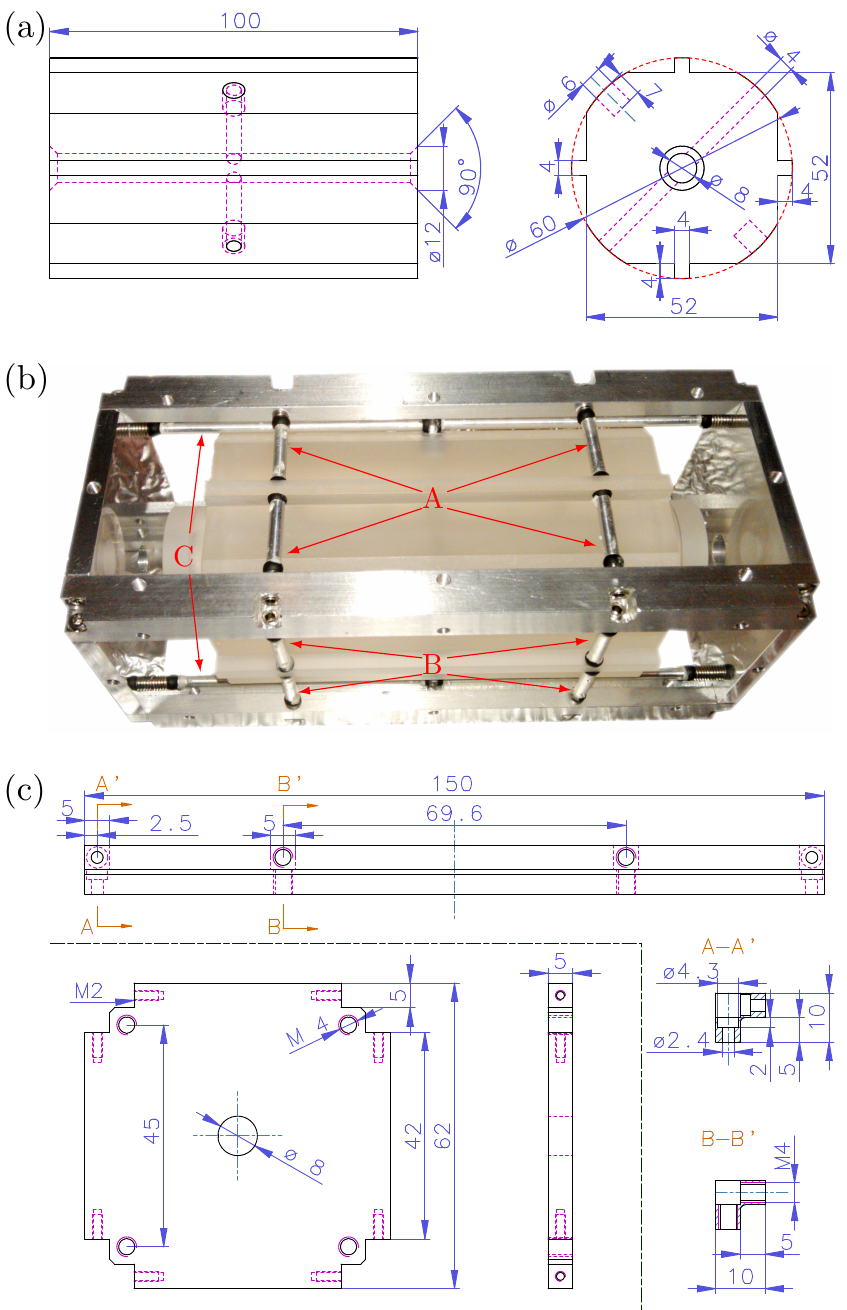}\protect\caption{\label{fig:drawing}(a) Technical drawing of the cavity spacer. (b)
Photo of the cavity in the supporting frame. (c) Simplified mechanical
drawing of the supporting frame. The unit is 1~mm.}
\end{figure}

\subsection{Design of the reference cavity}

The design of the cavity spacer and support frame are shown in Fig.~\ref{fig:drawing}.
The dimensions of the approximately cylindrical cavity spacer are
100~mm\,$\times$\,60~mm (length$\times$diameter). The detailed
design of the cavity spacer is shown in Fig.~\ref{fig:drawing}~(a).
The spacer is made from standard grade ULE glass. On each end of the
spacer a fused silica mirror (ATFilms, Boulder %
\footnote{This and other information on manufacturers is provided for technical
communication purposes only and does not constitute an endorsement
by the University of D\"{u}sseldorf or PTB.%
}) is optically contacted. The purpose of using fused silica mirrors
instead of ULE mirrors is to reduce the thermal noise \cite{numata2004thermalnoise,kessler2012thermal}
to a calculated instability flicker floor of $\sigma_{y}=4\times10^{-16}$.
The radii of curvature of the mirrors are infinity and 1~m, respectively.
The mirrors are \change[1~inch]{25.4~mm} in diameter and \change[1/4~inch]{6~mm} thick and are high-reflectivity-coated
for 1064~nm, leading to a measured cavity linewdith of 3.7 kHz and
finesse of 400~000. The central 6~mm diameter area of the back side
is anti-reflection-coated for the same wavelength. On the back sides
of the cavity mirrors, ULE rings (inner/outer diameter 9/25.4~mm,
thickness \change[1/4~inch]{6~mm}) are optically contacted to compensate the thermal
expansion caused by the fused silica mirrors \cite{legero2010tuningthe}. 

The mounting of the cavity is based on the idea to separate the forces acting
on the cavity along the optical axis and perpendicular to this direction. The
resulting forces are then applied in the corresponding symmetry plane of the
cavity. Thus, due to symmetry, to first order
in the accelerations $a$\change{,} \change[no change of ]{}the resonator length, i.e.
the distance between the mirror centers\change{,} \change[is expected]{will not change}
\cite{uwe2012frequenzstabilisierungsvorrichtung}. 

The mounting structure of the cavity is shown in Fig.~\ref{fig:drawing}~(b),
The cavity is supported by squeezing 20 short posts with diameters
of 3~mm pairwise on 10 points on the cavity spacer. These 20 posts
are divided into 3 groups A, B, and C in Fig.~\ref{fig:drawing}~(b).
The posts in a given group are parallel to each other. Each group
restricts the movement of the cavity in the direction parallel to
the posts' axes while it allows motion of the cavity in the two directions
perpendicular to the posts' axes. For motional constraint in the two
directions orthogonal to the cavity axis (groups A and B), 4\change{-}post
pairs are used for each direction. 2\change{-}post pairs are used in order
to limit the movement of the cavity along its axis (group C). The
distance along the cavity axis between the supporting points in groups
B and C are 69.6~mm and they are located symmetrically with respect
to the midplane of the cavity which is perpendicular to its axis.
The supporting points of group C are on this symmetry plane of the
cavity and act on invar pins that are glued into the ULE pacer. The
positions of the supporting points were optimized by using finite
element simulation \cite{millo2009ultrastable,chen2006vibrationinduced}.
M4 set screws on the frame are used to press the posts to the cavity.
Viton balls (diameter 4 mm) are inserted between the screws and the
posts to reduce the tangential restoring force of the posts. Viton
balls cut in half are inserted between the cavity and the posts so
as to increase friction and further decouple the cavity from the supporting
frame. The mechanical drawing of the supporting frame is shown in
Fig.~\ref{fig:drawing} (c).

\begin{figure}
\includegraphics[bb=60bp 20bp 720bp 520bp,clip,width=0.9\columnwidth]{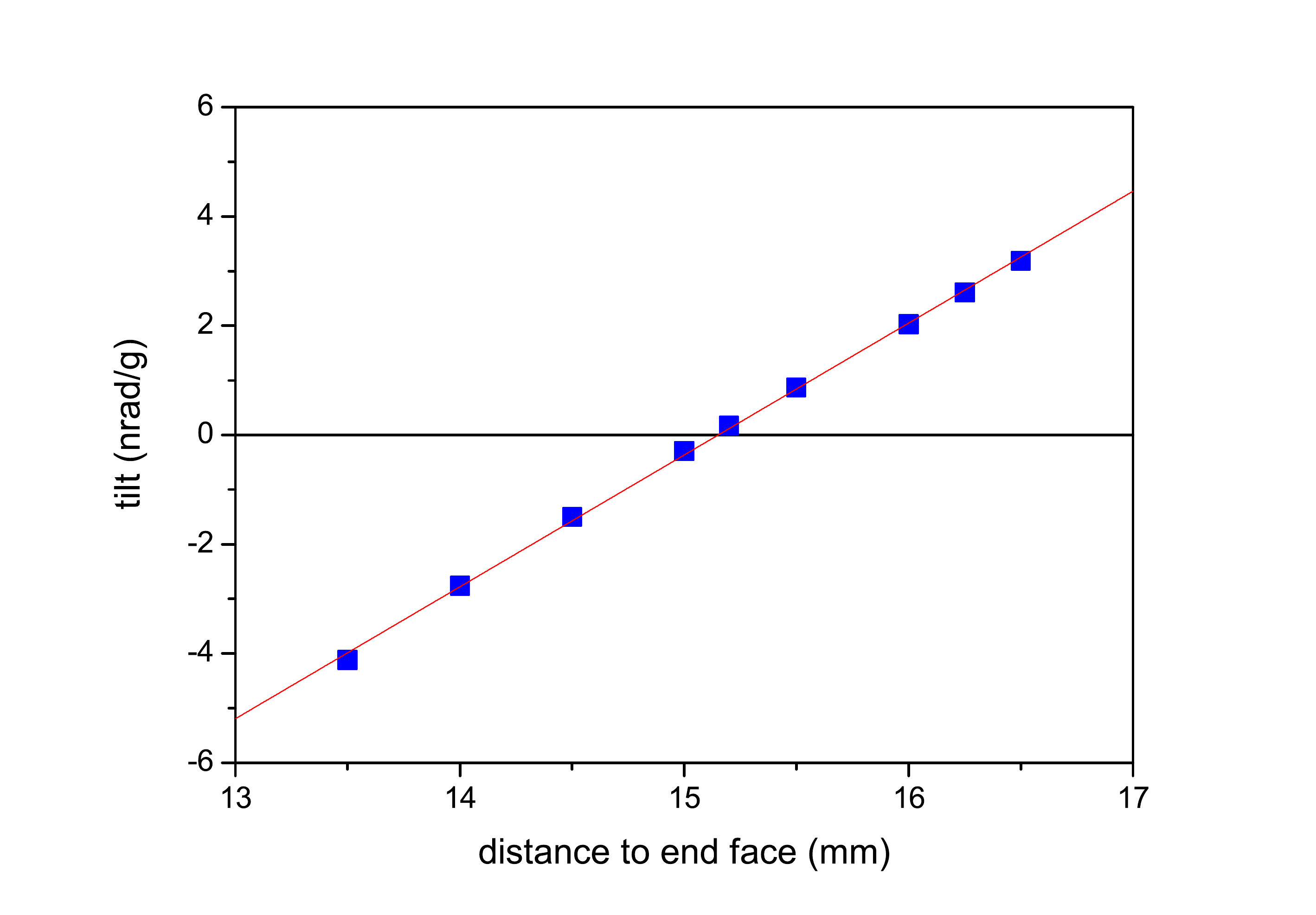}\protect\caption{\label{fig:tilt}Tilt of a single end mirror of the cavity as a function
of the mounting position under the acceleration $a=1\, g$ perpendicular
to the cavity axis.}
\end{figure}

\begin{figure}
\includegraphics[width=0.9\columnwidth]{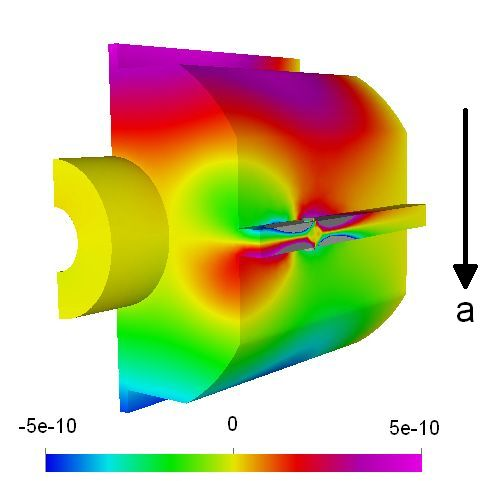}\protect\caption{\label{fig:bend}FEM calculation of deformation in the direction parallel
to the optical axis under an acceleration of $1\, g$. The color code
indicates the displacement along the cavity axis (in the unit of meter)
and for symmetry reason only one quarter of the cavity was simulated.}
\end{figure}

Effects still to consider are deviations from symmetry, e.g. from
an offset of the mode axis from the geometrical symmetry axis in combination
with acceleration induced bending of the spacer. This bending can
be minimized by appropriate positions of the mounting. This was simulated
by FEM models \cite{elmer} (Fig. \ref{fig:tilt}) to identify the
mounting points with the smallest bending (see Fig. \ref{fig:bend}).
Assuming a 1~mm deviation of the resonator mode from the cavity's
geometric symmetry axis, a deviation of the mounting position of 0.5~mm
from the optimum position would lead to a sensitivity of $\Delta L/(a\, L)\approx10^{-11}$/$g$. Deformation
from squeezing the rim at the mounting positions can also couple accelerations
to length changes. The FEM simulations indicate a value of $\Delta L/L\approx10^{-11}$
for a squeezing force of 1 N. To provide reliable holding under a design
acceleration of $a_{\rm max}=20\, g$, the squeezing force of the mounting
needs to be sufficiently large. For a perfectly linear elastic material,
the squeezing force needs to balance the maximum reaction force. In
our case this would require, for each of the four points in each direction,
a squeezing force of 15~N. However, the large-range behavior of viton
balls is strongly nonlinear, with the force increasing to the power
3/2 with approach \cite{tat91,tat91a}. This reduces the required
squeezing force by 30\% to 10~N. Implementing a higher nonlinearity,
e.g. by optimized shape of the viton parts, the squeezing force and
related deformation can even be further reduced.

Imperfections on the mount can lead to forces tangential to the mounting
plane. On the rim opposite axial forces of 1~N on each pad result
in $\Delta x=6.8\times10^{-10}$~m or $\Delta L/L=13.6\times10^{-9}$
and radial forces of 1~N at each pad lead to $\Delta x=2.19\times10^{-10}$~m
or $\Delta L/L=4.38\times10^{-9}$ via the Poisson effect. With the
cavity mass of 600~g at $1\, g$ acceleration, a total normal force
of 6~N needs to be taken up by the supports in group A or B, so these
tangential forces need to be less than $10^{-3}$ of the normal force
for a total sensitivity less than $10^{-12}$/$g$. 

\begin{figure}
\includegraphics[width=8cm]{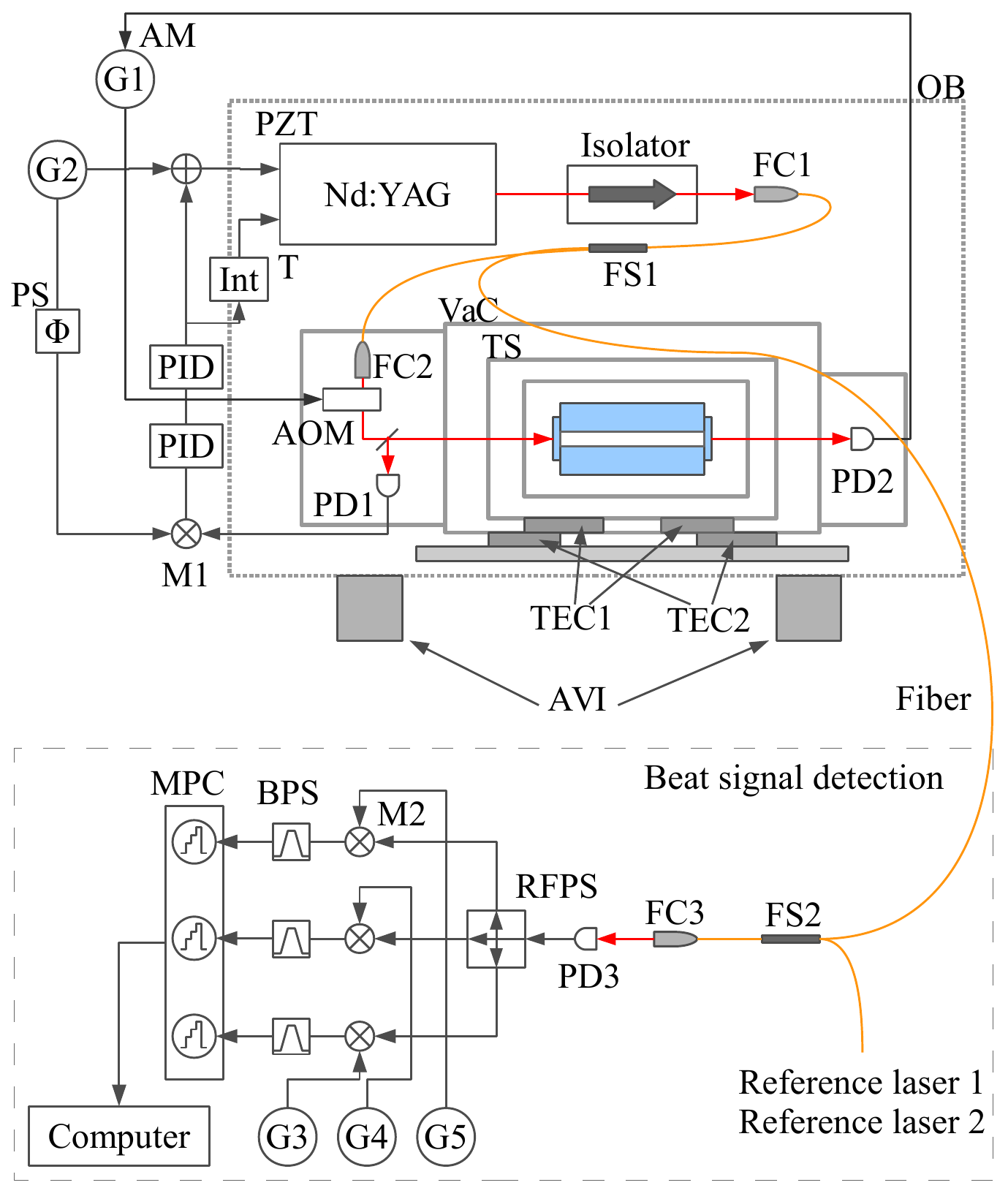}

\protect\caption{\label{fig:schematic}Schematic of the system. The details of the
compact optical setup is shown in Fig.~\ref{fig:overview}~(b).
The setup with the abbreviations is described in the main text. }
\end{figure}

\begin{figure}
\includegraphics[width=8cm]{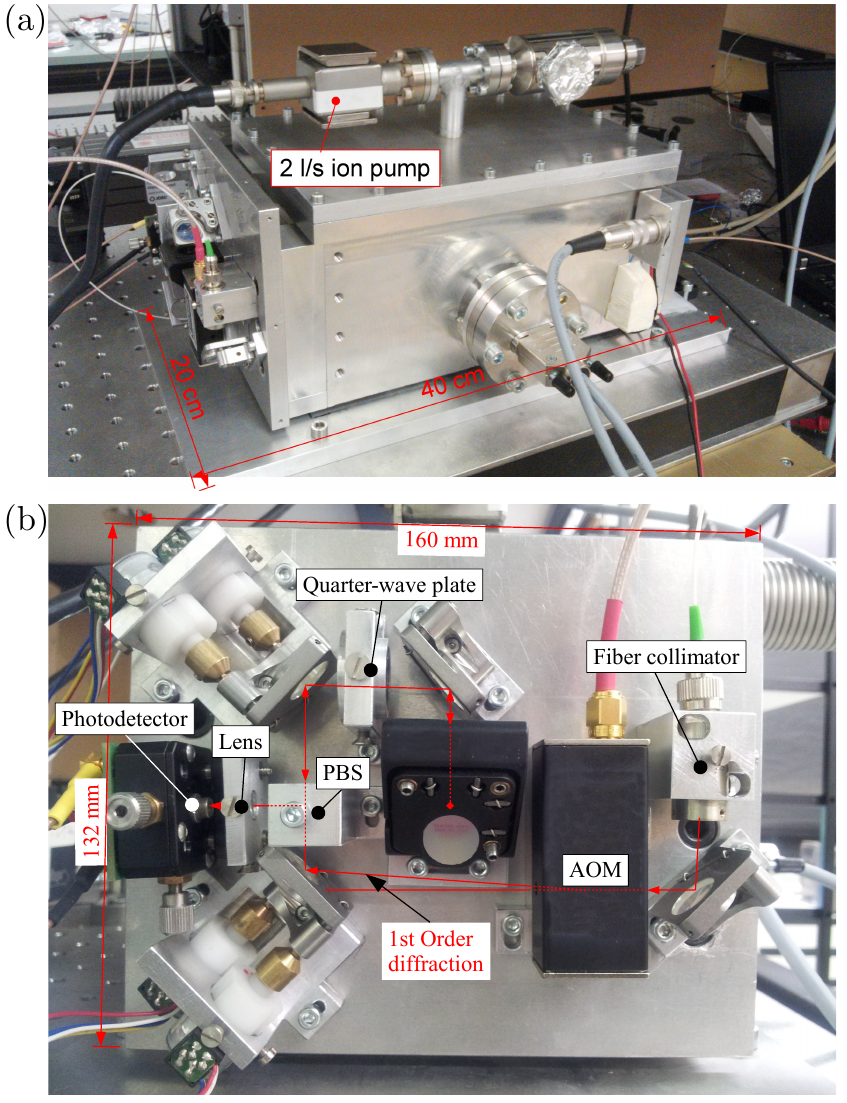}

\protect\caption{\label{fig:overview}(a) Overview of the system, with dimensions.
(b) the compact optical setup for coupling the laser beam to the cavity
and obtaining an error signal.}
\end{figure}

\subsection{Cavity subsystem description}

The cavity is operated in a vacuum chamber (VaC) inside two thermal
shields. The inner shield is formed by four polished aluminum side
covers (4~mm thick, not shown in the figures) that are fixed to the
support frame. This assembly is mounted inside the outer thermal shield
(TS) by screws and thermally isolating ceramic washers. This outer
thermal shield is mounted on 4 thermoelectric cooling elements (TEC)
on the bottom of the vacuum chamber using indium foils to insure good
thermal contact between the TECs and the thermal shield and the chamber.
The TECs allow to actively stabilize the temperature of the outer
thermal shield to the zero coefficient-of-thermal-expansion (CTE)
temperature of the cavity, which was determined to be near 0~$^{\circ}$C
by measuring the frequency of the cavity when set to different temperatures.
Remaining temperature fluctuations of the outer shield are additionally
filtered by the inner heat shield, leading to an overall second-order
low-pass behavior between the outer shield and the cavity. On long
time scales its behavior is described by a time constant of 5.5~hours,
which was determined by measuring the the laser frequency change after
a step of the outer shield temperature. To further improve the temperature
stability, also the temperature of the vacuum chamber is actively
stabilized at 20~$^{\circ}$C. All temperature sensors are standard
10~k$\Omega$ thermistors and PID control circuits are used to stabilize
the temperatures. The in-loop temperature stability at the outer thermal
shield is about 1~mK. The vacuum in the chamber is kept to $1\times10^{-7}$~mbar
by a miniature 2 l/s ion pump (\change{Gamma Vacuum}).

\subsection{The optical system}

A schematic of the complete system is shown in Fig.~\ref{fig:schematic}.
A Nd:YAG laser (Innolight, Mephisto S) is frequency-modulated at 3~MHz
(G2) via the piezo-actuator (PZT) of the laser. After an optical isolator, the
laser beam is coupled to a single-mode fiber \change{by an angled physical
contact (APC) fiber collimator (FC, FC1)}. Using a fiber splitter
(FS1) one part of the laser wave is sent to a compact optical breadboard,
which is attached to the vacuum chamber Fig.~\ref{fig:overview}~(a)
and further coupled into the cavity. The detailed design of the compact
breadboard is shown in Fig.~\ref{fig:overview}~(b). The laser beam exiting from
\change[the FC/APC collimator ]{FC2} is diffracted by the acousto-optic
modulator (AOM, Crystal Technology 3200-147). The first-order diffracted
wave is coupled into the cavity. The polarizing beam splitter (PBS)
and the quarter-wave plate are used to pick up the reflection from
the cavity. The lens and the photo-detector PD1, mounted on a translation
stage are used to detect the reflection from the cavity. The electric
signal from the PD1 detector is frequency-mixed (by M1, \change{Mini-Circuits} ZAD-8)
with the 3~MHz signal, phase-shifted (PS) with respect to the laser
modulation signal to generate the error signal. It is filtered by
two cascaded proportional-integral-derivative (PID) circuits with
P-I corner frequencies of approximately 3 kHz and 10 kHz and P-D corner
frequency of approximately 20~kHz and 30 kHz. The output of the locking
circuits combined with the 3~MHz modulation by using a bias-T (\change{Mini-Circuits} ZFBT-4R2GW) and fed back to the PZT modulation input of the laser.
The output from the second PID is also further integrated (INT) and
sent to the temperature control modulation input (T) of the laser
for long-term stabilization of the laser \cite{chen2012locking}.
The transmission of the cavity is detected by another detector (PD2)
attached to the other side of the vacuum chamber. This signal is used
to stabilize the laser power circulating inside the cavity through
feedback to the amplitude modulation input (AM) of the AOM driver
(G1). The sensitivity of the frequency on the transmit power is about
100~Hz ($3.5\times10^{-13}$) per 1~\%. 

Four stepping motors (\change{Nanotec }SPG1518M0504-102\change[, Nanotec]{}) are attached to two
of the mirror mounts (\change[9876, ]{}New Focus\change{{ 9876}}), which may be used to optimize
the beam coupling to the cavity when necessary and when there is no
manual access to the optical setup, e.g., when during operation in
space or at a location on Earth distant from the laboratory. The motors
can therefore be omitted when the system is operated in a laboratory
with manual access. During initial alignment of the small breadboard,
a low-cost CCD camera is set up in transmission through the cavity,
in order to identify the cavity modes. 

The laser and the optical system are attached to a standard commercial
optical breadboard (OB), and placed on an active-vibration-isolation
system (AVI, \change[AVI-400, ]{}Table-Stable\change{{ AVI-400}}). This arrangement is used for
testing and characterizing the system, and for operation on Earth.
In space, the latter two components would be absent. 

The overview of the system is shown in Fig.~\ref{fig:overview} (a).
The outer dimensions of the system, without the laser, optical breadboard,
and vibration isolation, are about $400\times200\times300$~mm\textsuperscript{3}
(length$\times$width$\times$height). The mass is approximately 10~kg.

\begin{figure}
\includegraphics[width=8cm]{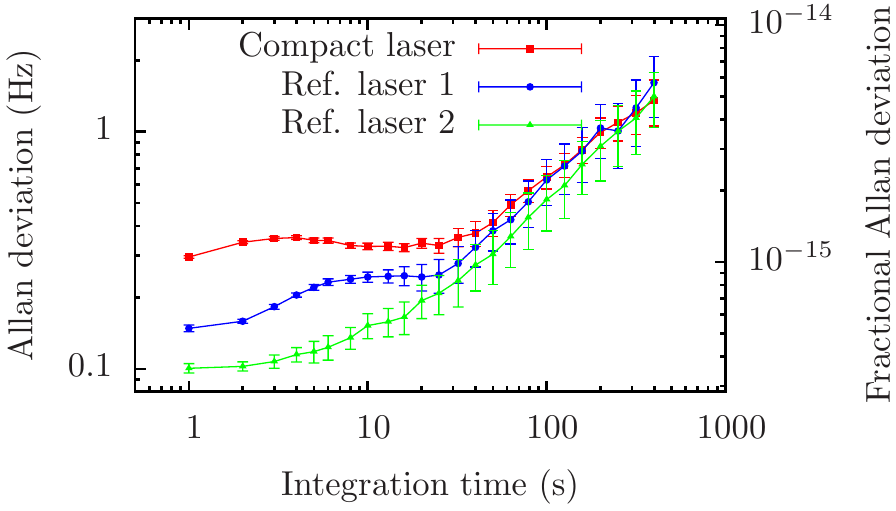}

\protect\caption{Allan deviation of the frequency instability of each laser derived
with the three-cornered-hat method, when they are frequency-stabilized
to independent ULE cavities. It is the average of 10 pieces of Allan
deviation calculated from 2000~s long beat frequencies with linear
drift removed.}

\label{fig:allan}
\end{figure}

\begin{figure*}
\includegraphics[height=4cm]{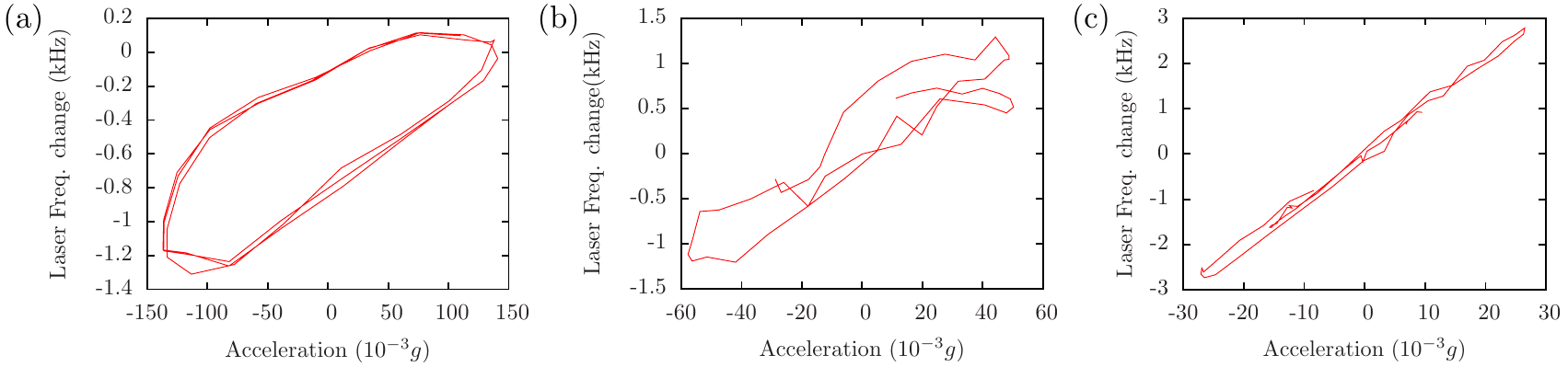}

\protect\caption{\label{fig:vibration}Relation between the system acceleration and
the resulting laser frequency change. (a) the system is shaken in
vertical direction, (b) the system is shaken in horizontal direction,
and perpendicular to the cavity axis, (c) the system is shaken along
the cavity axis. }
\end{figure*}

\begin{figure}
\includegraphics[width=8cm]{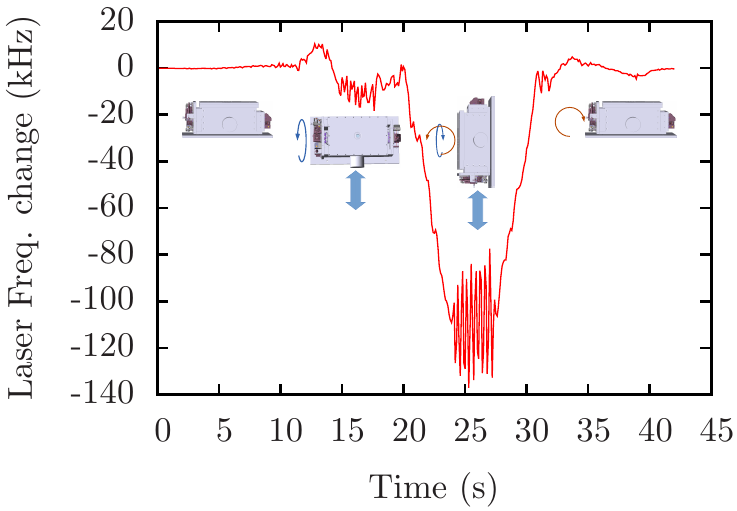}

\protect\caption{\label{fig:rotate}Laser frequency change while the system is manually
reoriented and shaken. The cavity housing images show the orientation
and acceleration directions in the different phases of the test. The
thin arrows show the rotating direction of the chamber. }
\end{figure}

\subsection{Characterization method}

The frequency stability of the compact laser system is characterized
by beating with two independent ultrastable Nd:YAG lasers whose frequencies
are locked to two reference cavities with lengths of 30~cm. The frequency
instability of each individual laser is then obtained by the three-cornered-hat
method \cite{gray1974amethod}. The laser beam exiting FS1 (Fig. \ref{fig:schematic})
is combined in the fiber splitter FS2 with the two beams from the
two reference lasers. \change[The three beat signals between the three laser
waves are detected by a single, fast photo-detector PD3 (EOT ET-3000A).]{The combined beam exiting from FC3 is guided to a fast photo-detector PD3 (EOT ET-3000A), for detecting the three beat signals between the three laser
waves.}
The beat signals (around 70, 480, and 550~MHz) were split into
three ways \change[(by ZF3RSC-542-S+, mini-circuits)]{by RFPS (RF power splitter, Mini-Circuits ZF3RSC-542-S+)} and down-mixed by three
independent synthesizers \change{G3, G4, and G5 }(HP 8656B, and phase-locked to the same 10~MHz
reference) to approximately 10~MHz \change{using mixers M2}. Each signal was bandpass filtered
at 10~MHz \change[(by BBP-10.7+, mini-circuits)]{by BPS (bandpass filter, Mini-Circuits BBP-10.7+)} and sent to a four-channel
dead-time-free phase comparator \change{MPC }(\change[K+K FXE, ]{}K+K Messtechnik GmbH\change{{ K+K FXE}}). The
frequencies are recorded by a computer and analyzed to obtain the
Allan deviations.

\subsection{Characterization results}

We operate the cavity with 200 $\mu$W entering the vacuum chamber.
The coupling efficiency of the laser wave to the cavity is approximately
40~\%. The bandwidth of the lock system is about 70~kHz. The system
works very reliably and essentially never fell out of lock during
several months of operation. The frequency instabilities of the compact
laser and the two reference lasers, obtained simultaneously via the
three-cornered-hat method are shown in Fig.~\ref{fig:allan}. The
lowest instability of the compact laser is $1\times10^{-15}$ for
integration times of 1 to 10~s. It remains below $3\times10^{-15}$
for integration times up to 100 s. This performance satisfies the
specification of the STE-QUEST mission \cite{stequest} which calls
for this level for integration times up to 100~s.

The acceleration sensitivity coefficients of the cavity are determined
by recording acceleration level and beat frequency with respect to
one reference laser, while shaking the system (optical system plus
optical breadboard placed on AVI, which is supported by springs) sequentially
in the three directions. The acceleration level is measured by an
acceleration sensor (Wilcoxon research, model 731A) attached to the
optical breadboard. The beat frequency is converted to a voltage by a
frequency-to-voltage converter (home-built, based on a \change{
frequency-to-voltage converter chip, Texas Instruments }VFC32KP).
The output from the acceleration sensor and the frequency signal are
simultaneously recorded by a digital-analog data acquisition card. 

An acceleration along the vertical direction was excited by a voice
coil actuator (home-built) attached to the optical table and to the
optical breadboard and driven at a frequency close to the resonance
of the overall system. The acceleration along the two horizontal directions
were excited by manually hitting the optical breadboard on the side.
From the test, shown in Fig.~\ref{fig:vibration}, we obtained acceleration
sensitivities along the vertical, the horizontal across-cavity, and
the axial directions of $1.7\times10^{-11}/g$, $8.0\times10^{-11}/g$,
and $3.9\times10^{-10}/g$, respectively.

For testing the robustness of the system, the system was carried,
rotated and shaken by two persons holding it in their hands. The laser
frequency remains stably locked to the cavity. The laser frequency
change during this procedure is shown in Fig.~\ref{fig:rotate}.
The frequency change is about 20~kHz when the cavity is rotated along
the cavity axis by 90 degrees (starting at 12~s in the figure), and
it is about 100~kHz when the cavity is oriented from the horizontal
into the vertical direction (rotation occurs around the horizontal
cross-axis, at 40~s). These frequency shifts are consistent with
the sensitivities observed in Fig.~\ref{fig:vibration}. This test
demonstrates that the cavity can be operated vertically or horizontally,
and that the mounting concept is quite robust.

Because the cavity and the setup may experience a typical temperature
variation from -20$^{\circ}$C to +50$^{\circ}$C during storage and
transportation to space, a test of the robustness of the optical setup
with respect to temperature variation was also done. The optical setup
(without laser and with temperature stabilization switched off) was
put in a refrigerator, cooled down to -30$^{\circ}$C, then warmed
up to 50$^{\circ}$C with the TEC attached to the vacuum chamber,
and finally back to room temperature. The coupling efficiency of the
laser beam into the cavity changed from 40~\% to 15~\% during this
temperature cycle. This means that the alignment was not ``lost'';
it was brought back to 40~\% by adjusting just one mirror mount.
This was done manually, but, as described above, can in the future
also be done by remote control or an automatic routine, using the
motorized mirror mounts in the small optical breadboard.

\subsection{Conclusion}

In summary, we reported a robust mounting structure for high-finesse
cavities, which represents a baseline design for a space unit. In
addition, since it does not at present include any particularly sophisticated
or expensive elements, the design is also suitable for laboratory
use, and for non-laboratory, terrestrial environments. Applications
are e.g. for a transportable optical clock or for a transportable,
optically stabilized frequency comb. As an example of a useful application,
in our laboratory we routinely use this system as an ultra-stable
optical local oscillator to which a fiber frequency comb is locked
that measures several other ultra-stable optical frequencies. 
\begin{acknowledgments}
This work was funded by the Bundesministerium f\"ur Wirtschaft und
Technologie (Germany) under project no. 50OY1201 and the European
Metrology Research Programme (EMRP) under IND14. The EMRP is jointly
funded by the EMRP participating countries within EURAMET and the
European Union.
\end{acknowledgments}

\end{document}